\documentclass[12pt,preprint]{aastex}

\shorttitle{HD~69686}
\shortauthors{Huang et al.}

\begin{document}

\received{}
\accepted{}

\title{HD~69686: A Mysterious High Velocity B Star}

\author{Wenjin Huang}

\affil{Department of Astronomy \\
University of Washington, Box 351580, Seattle, WA 98195-1580;\\
hwenjin@astro.washington.edu}

\author{D. R. Gies}

\affil{Center for High Angular Resolution Astronomy\\
Department of Physics and Astronomy \\
Georgia State University, P. O. Box 4106, Atlanta, GA  30302-4106;\\
gies@chara.gsu.edu}

\author{M. V. McSwain}

\affil{Department of Physics, Lehigh University\\
16 Memorial Drive East, Bethlehem, PA  18015;\\
mcswain@lehigh.edu}


\slugcomment{Submitted to ApJ}



\begin{abstract}
We report on the discovery of a high velocity
B star, HD~69686.  We estimate its space velocity, distance,
surface temperature, gravity, and age.
With these data, we are able to reconstruct the trajectory of the star
and to trace it back to its birthplace.  We use evolutionary tracks for
single stars to estimate that HD~69686 was born
73 Myr ago in the outer part of our Galaxy ($r \sim 12$ kpc) at a
position well below the Galactic plane ($z \sim -1.8$ kpc), a very
unusual birthplace for a B star.  Along the star's projected path
in the sky, we also find about 12 other stars having similar proper
motions, and their photometry data suggest that they are located at the
same distance as HD~69686 and probably have the same age.  We speculate
on the origin of this group by star formation in a high velocity
cloud or as a Galactic merger fragment.
\end{abstract}


\setcounter{footnote}{0}
\section{Introduction}                              
The objects in the solar neighborhood with high peculiar velocities
can always be placed into one of two categories: the disk-origin or
non-disk origin.  The normal, young, disk stars in the solar neighborhood
share the same circular motion with the Sun, and have small velocity
residuals in a Gaussian distribution with a standard deviation
of 10-16 km~s$^{-1}$ \citep[and references therein]{gie86}. 
The disk-origin high velocity stars are the objects whose peculiar
velocities are far beyond the velocity dispersion of normal disk stars.
They are often called runaway stars because the high peculiar velocities of
these stars are thought to be caused by two possible mechanisms:
1) supernova explosion in a binary system \citep{zwi57,bla61}; or 2)
strong gravitational interaction in a multiple star system that often occurs
in a dynamically chaotic, young, dense stellar cluster environment
\citep{pov67,leo88,gab08}.  Previous studies
of runaway stars show that both mechanisms generate runaway
objects, but an early systematic survey by \citet{gie86} suggests that the
second mechanism is dominant.  On the other hand, most of the non-disk-origin,
high velocity objects in the solar neighborhood are thought to belong to the old
halo population.  The halo stars have high peculiar velocities because
they do not follow the circular motion of the normal disk stars but move
in their own very diverse orbits.  Because of the significant difference
in the ages of high velocity stars associated with these two origins, 
the runaway, disk origin is generally assumed for young objects like 
high velocity B stars.  However, in this paper, we discuss the unusual 
case of a nearby, high velocity B-star, HD~69686, and we argue that
it may have been born far from the Galactic disk. 

We are currently making a spectroscopic survey of field B stars to 
study the evolution of stellar rotation.  One of our targets, 
HD~69686, was selected because it is fairly
bright ($V\sim7.1$) and is classified as a B8 star in the SAO star catalog.
In this paper, we present what we have learned about HD~69686.  
At the time of writing, the target's listing in the SIMBAD 
database includes only very basic astrometry and photometry data 
and no published references.  After measuring radial velocities
of our program stars, HD~69686 was immediately singled out by having
the largest radial velocity, 148 km~s$^{-1}$, among our sample B stars.
Only after this, we noticed that this star also has quite a large proper
motion ($\mu_\alpha \cos \delta = -86.17\pm0.67$ mas~yr$^{-1}$, 
$\mu_\delta = 7.21\pm0.42$ mas~yr$^{-1}$, from the newly released
{\it Hipparcos} re-measurements by \citealt{van07}).  Thus, HD~69686
is a previously unknown high velocity B star.  How old is this star now?
Is it a runaway star? Where did it come from?  Is it a binary system?
These are the kind of questions that we want to answer.

In the next section, we briefly describe the instrumental setup and how
we obtained the final spectra of HD~69686. In \S3, we
present the details of the procedure we used to determine the key parameters, such
as the space velocity, the effective temperature, gravity, and age
of the star.  In \S4, we use a modern Galactic potential model to reconstruct
the full trajectory of HD~69686 that helps us to locate its birthplace and leads
us to suggest that HD~69686 might not be a runaway star as we thought previously.
In the final section, we provide additional evidence implying that HD~69686 may
been born in the Galactic halo.


\section{Observation}                       

The spectra of HD~69686 were obtained on the 2.1-m telescope
at KPNO using the Goldcam spectrograph (with a $3072\times1024$
CCD detector, T3KC) on 2008 November 17 and 18.
The spectrograph grating (G47, 831 lines mm$^{-1}$) was
used in second order with a CuSO$_4$ blocking filter, 
and this arrangement provided spectrum
coverage of about 900\AA\ around the central wavelength of 4400\AA.  
The slit width was set at $1\farcs3$, leading to a resolving
power $R\sim 2400$ (FWHM$\sim$1.83 \AA, measured in the comparison spectra).
The integration time for HD~69686 was 55~s yielding 
spectra with S/N $\sim$ 280 in 
the continuum regions.  During the entire
observing run, we took comparison (HeNeAr lamp) exposures
for each of our program stars, including HD~69686, to
ensure accurate wavelength calibration of the spectra.
The accuracy of wavelength calibration is estimated to be around
6 km~s$^{-1}$ in velocity space by checking the wavelength fitting
residuals of the comparison spectra and by comparing multi-night
spectra of all our program stars that are not spectroscopic
binaries.  We obtained the final reduced spectra by going through
the standard IRAF\footnote{http://iraf.noao.edu/} CCD image reduction
(subtracted the bias level, divided the flat images, removed cosmic
rays, and fixed the bad pixels/columns) and the long-slit stellar
spectrum extraction procedures (traced and binned the spectrum,
calibrated wavelength).


\section{Stellar Parameters}       

The procedure to determine the physical parameters of HD~69686
is similar to that used in our previous studies of cluster
B stars \citep{hua06a,hua06b}. The $V \sin i$ value was derived
by fitting synthetic model profiles of \ion{He}{1}
$\lambda 4471$ and \ion{Mg}{2} $\lambda 4481$, using realistic
physical models of rotating stars (considering Roche geometry
and gravity darkening).  The details of this step are
described in \citet{hua06a}.  The best fit of \ion{He}{1} $\lambda 4471$
is illustrated in Figure~1.  With $V \sin i$ at hand, we
then determined both the effective temperature ($T_{\rm eff}$) and
gravity ($\log g$) of the star by fitting the H$\gamma$
profile (see details in \citealt{hua06b}).  The best fit of
the H$\gamma$ profile is shown in Figure~2.  By shifting the best fit
profiles in wavelength to match the observations we also obtained
an accurate radial velocity of each night's spectrum.  The radial
velocities were transformed into the heliocentric frame by removing
the orbital motion of the Earth (using the RVCORRECT function in IRAF),
and these velocities are given in Table~1.
We do not see any significant change in the radial velocity between
the two nights that is larger than the errors (6 km~s$^{-1}$). Even
if HD~69686 were a long period, spectroscopic binary system, any
orbital velocity variations will be small compared to its radial velocity.  
Thus, we adopted the mean of the radial velocities in Table~1 as the
systemic radial velocity of HD~69686 for later analysis (Table~2).
For a rotating star such as HD~69686, the derived gravity ($\log g$)
represents an average of gravity over its visible hemisphere.  It may
not be a good indicator of evolutionary status of the star because
the effective gravity on the equatorial region is lowered by
centrifugal force induced by stellar rotation.  Instead, we use
the gravity at the poles of the star ($\log g_{\rm polar}$) to
estimate the age of the star since the $\log g_{\rm polar}$ value is
not significantly influenced by rotation.  The method to estimate $\log g_{\rm polar}$
is described in \citet{hua06b} and has been applied to many of
our studies on the evolution of B stars \citep{hua08,mcs08}.  Based on
our model simulation results \citep{hua06b}, except for those extreme
cases where the stars spin at or very close to the breakup velocity,
the estimations of $\log g_{\rm polar}$ are quite accurate
(the statistical errors are 0.03 dex or less).  

\begin{figure}
\plotone{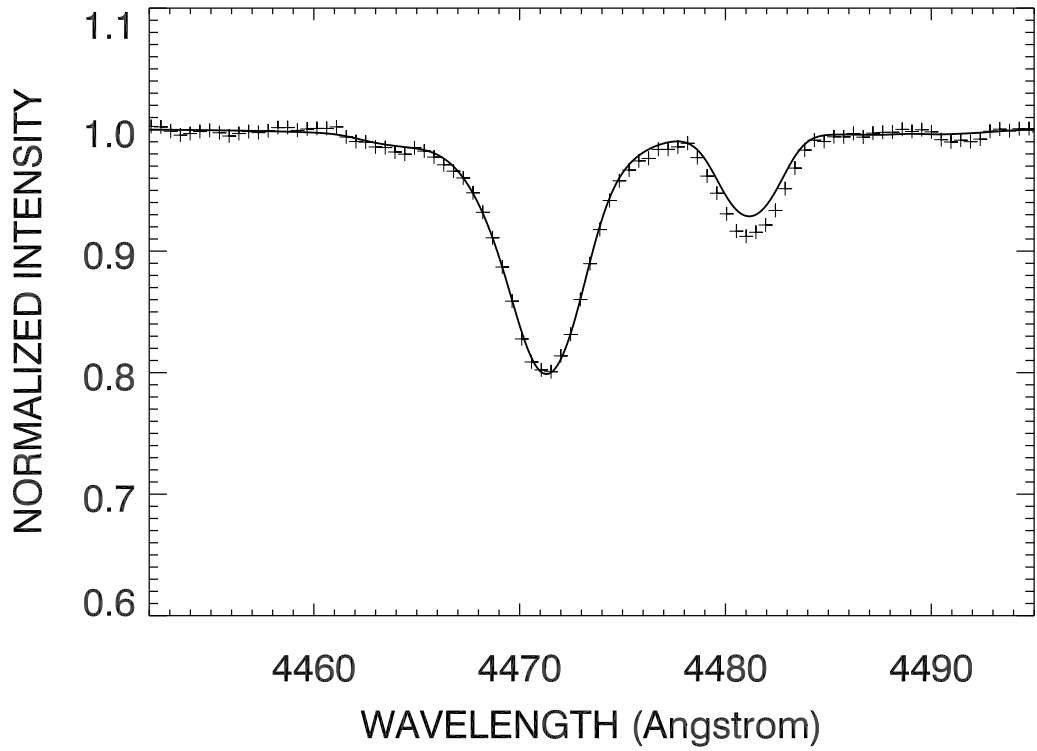}
\caption{The best fit solar abundance model (H:He = 0.9:1.0) of
\ion{He}{1} $\lambda 4471$ ({\it solid line}) compared to the observed
spectrum ({\it plus signs}).  The model parameters are:
$T_{\rm eff}=16200$ K, $\log g=3.92$, $V\sin i=141$ km~s$^{-1}$.}
\label{fig1}
\end{figure}

\begin{figure}
\plotone{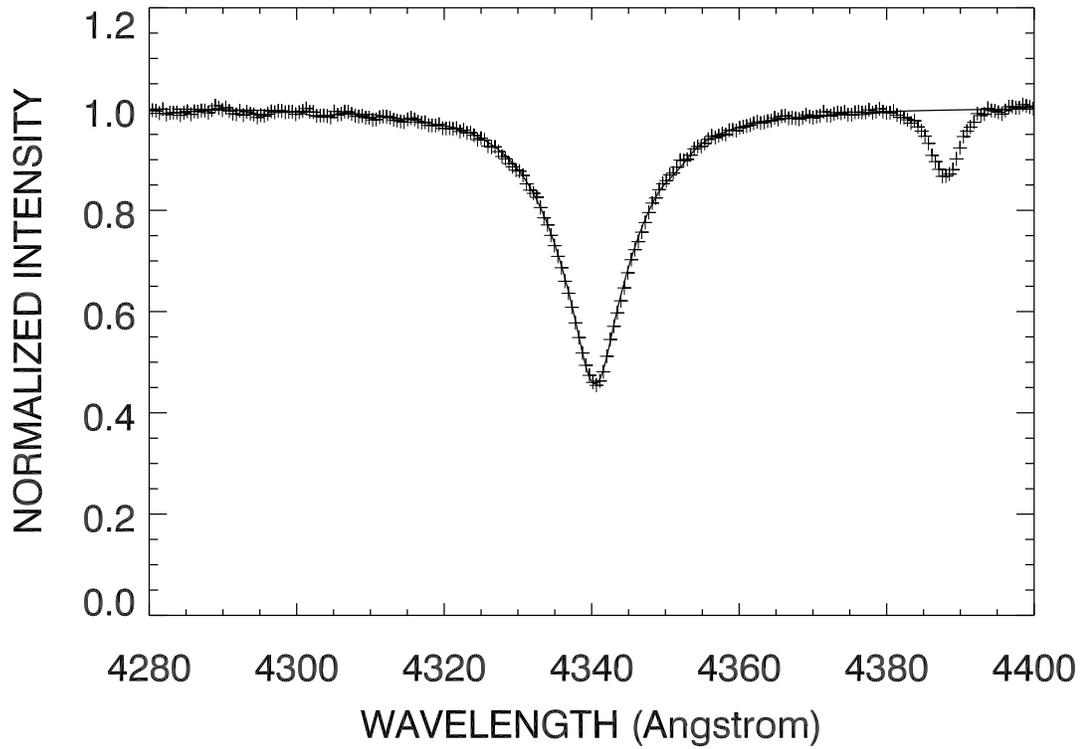}
\caption{The best fit solar abundance model (H:He = 0.9:1.0) of
the H$\gamma$ profile ({\it solid line}) compared to the observed
spectrum ({\it plus signs}).  The model parameters are:
$T_{\rm eff}=14500$ K, $\log g=3.91$, and $V \sin i=141$ km~s$^{-1}$.
Note: the \ion{He}{1} $\lambda 4387$ profile is not included in
the synthesized spectrum.}
\label{fig2}
\end{figure}

\placefigure{fig1}     
\placefigure{fig2}     
\placetable{tab1}      

We need one additional step to correct $T_{\rm eff}$ and $\log g$
derived from the H$\gamma$ fit.  We noticed that we had to use a
model with much higher temperature (16200K) to fit the \ion{He}{1}
$\lambda 4471$ line while our H$\gamma$ line fit suggests that the effective
temperature of HD~69686 should be around 14500K.  This discrepancy
implies that HD~69686 is a helium strong star.  When we use
models with solar abundances (H:He$=$0.9:0.1) to synthesize
the H$\gamma$ profile and to fit the observed spectrum of a He-peculiar star,
we get systematic errors in the derived parameters ($T_{\rm eff}$ and $\log g$).
The systematic errors can be corrected using the results given by \citet{hua08}
who used the same technique to determine
the photospheric parameters.  The correction procedure was done in an
iterative manner.  We first determine the He abundance using
the derived $T_{\rm eff}$ and $\log g$ values, then correct
$T_{\rm eff}$ and $\log g$ based on the He abundance according
to Table~3 of \citet{hua08}.  We then determine the He 
abundance again with the updated  $T_{\rm eff}$ and $\log g$ values,
and repeat the steps until convergence is reached.  At the end,
we obtained the final corrected $T_{\rm eff}$ and $\log g$ (listed
in Table~2) for HD~69686, with abundances typical of a He-strong star,
with a number ratio of H:He$=(0.82\pm 0.04):(0.18\mp 0.04)$.

The best available parallax and proper motion data of HD~69686 come from
the {\it Hipparcos} catalogue \citep{van07}.  However, the parallax
measurement, $2.35\pm0.58$ mas (1$\sigma$ range: 341$\sim$565 pcs), is
still too uncertain to provide us with an accurate tangential velocity 
for tracing the star's trajectory in space.
We have to find an alternative method to obtain a reliable 
distance of the star.  Our method is described as follows:
1) by comparing the derived $T_{\rm eff}$ and $\log g_{\rm polar}$
of HD~69686 with evolutionary tracks of stellar
models \citep{sch92}, we can estimate its stellar mass; 2) we then
can calculate the stellar radius from the mass and $\log g_{\rm polar}$;
3) from the radius and surface temperature, we can calculate the true
luminosity of the star (we considered a distorted shape of
a rotating star in this step, see details below);
4) with the true luminosity  and the $V$ magnitude of the star, we can
derive the distance for assumed values of extinction and bolometric correction.
The $B-V$ color index from the {\it Hipparcos/Tycho}
catalogue\footnote{http://archive.ast.cam.ac.uk/hipp/hipparcos.html}
(ESA 1997) is $-$0.15$\pm$0.01 which is basically no different from
the zero-reddening color of a star with temperature of 14,800~K.  Thus,
we can safely set $A_V=0$ in the last step.  The bolometric correction data we
used here are from \citet{bal94}.  Finally, we estimated the age of the star from
its mass and polar gravity by comparing with the non-rotating theoretical
models by \citet{sch92}.  The whole procedure was also repeated  with other denser
model grids by \citet{lej01},\citet{dem04}, and \citet{mar08}, and we found
no noticeable difference in the derived parameters between these
model grids.  All the derived parameters are listed in Table~2.  

\begin{figure}
\plotone{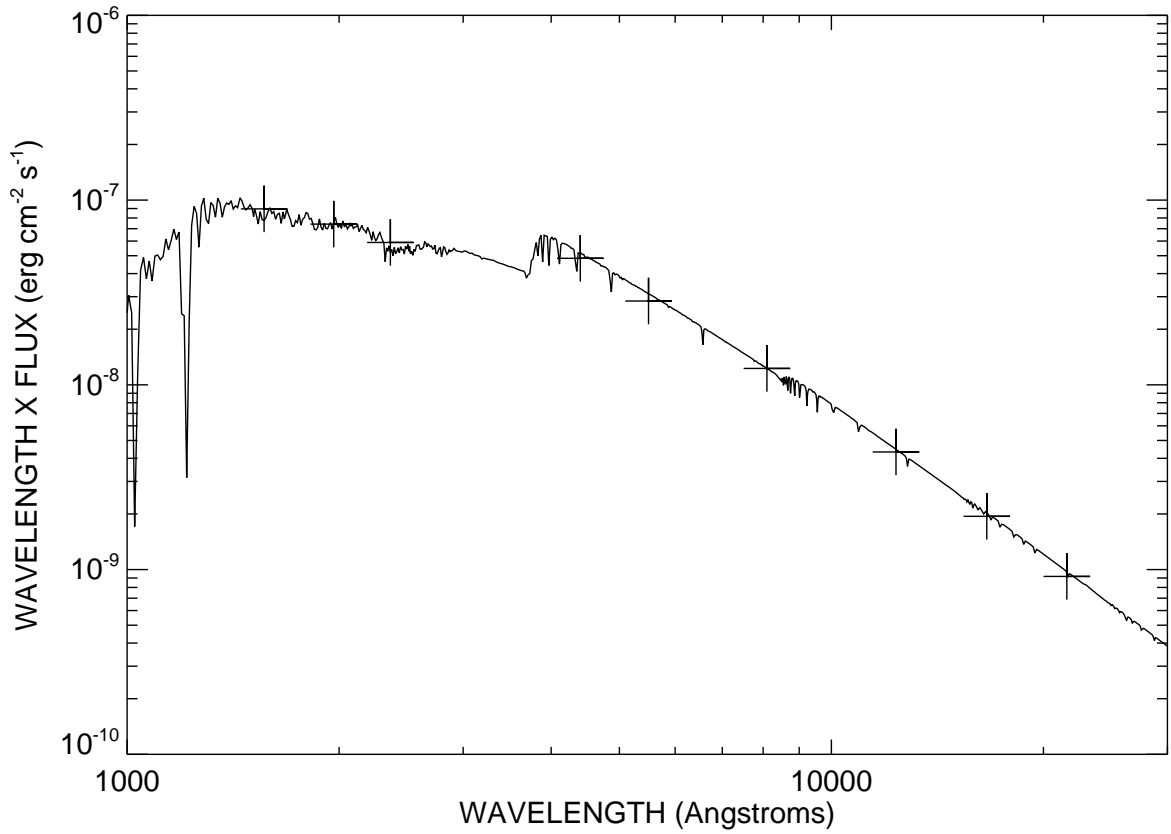}
\caption{The theoretical unreddened SED based on a Kurucz
model \citep{cas03} with $T_{\rm eff}=14760$, $\log g=3.93$, and an angular
diameter of 0.085$\pm$0.004 mas (limb darkened).  The optical/IR
(transformed from the $B$, $V$, $I_C$ and $JHK_S$ magnitudes) and UV flux
measurements are plotted as plus signs. }
\label{fig3}
\end{figure}

\placefigure{fig3}     
\placetable{tab2}      

\citet{heg00,mey00} and, more recently, \citet{eks08} (EKS models,
hereafter) have predicted that, due to rotationally induced mixing, fast
rotating stars may evolve on MS tracks quite differently from non-rotating
stars.  Fresh hydrogen can be brought down to the core due to rotationally
induced mixing and make rotating stars have a longer MS lifetime.  
To investigate the difference in ages between rotating and non-rotating
stars, we calculated the ages of the rotating models with various spin rates
at a fixed evolutionary status (at $\log g_{\rm polar} = 4.04$).  We choose the
3 $M_\odot$ model (the rotating model with mass closest to HD 69686) from EKS
models.  The result is given in Table~3.  As expected, fast rotating
stars are evolving more slowly than slow- or non-rotating stars with the
same mass in term of surface gravity at poles.  As shown in Table~3, a very
fast rotating star with $\Omega/\Omega_{\rm crit}=0.9$ probably takes
20\% more time than a non-rotating star to reach $\log g_{\rm polar} = 4.04$.

\placetable{tab3}      

It is worthwhile noting that HD~69686 is a rotating star with a
moderate $V \sin i$ value ($=141$ km~s$^{-1}$).  Because of
the centrifugal force on its surface, the projected disk
of HD~69686 on the sky is not perfectly round, and the projected
angular area is expected to be slightly larger than $\pi R_{\rm polar}^2/D^2$
($D$ is the distance) by a factor $f$, which depends on the
inclination angle $i$.  Because
we do not know the exact value of the inclination angle, we
can only estimate it by averaging all possible values of $i$.
The range of $i$ is between $\arcsin(V \sin i /V_{\rm crit})$
and $\pi/2$, where $V_{\rm crit}$ is the breakup velocity of
the star ($V_{\rm crit}=411$ km~s$^{-1}$ for HD~69686).  We
found the statistical mean of the acceptable 
inclination range is $i = 60^\circ$.  If we assume that 
HD~69686 (with $M=4.42 M_\odot$, $\log g_{\rm polar}=4.04$,
and, therefore, $R_{\rm polar}=3.32 R_\odot$) 
has an inclination of $i=60^\circ$, then 
we obtain $R_{\rm eq}=3.51 R_\odot$ and $f=1.07$ from calculations
based on simple Roche geometry.  With estimates in hand for the 
effective temperature and gravity, we can compare the observed
and model spectral energy distributions (SED) of the star
to determine its angular size, and then use our estimate of
its physical size to derive a second determination of distance.
In Figure~3, we plot the optical/IR flux points that are converted
from the $B$, $V$, $I_C$, and $JHK_S$ magnitudes of the star and from
the UV data from the ESRO TD$-1$ satellite \citep{tho78}.  The fit between
the calculated SED ($A_V=0$) and photometry data is very good, and
the normalization factor yields an equivalent, limb darkened angular
diameter of $0.085 \pm 0.004$ mas.  Using the projection factor $f$
and polar radius given above, we arrive at a distance
of $377 \pm 16$ pc, which is fully consistent with the 
spectroscopic distance derived earlier.  

We also apply our H$\gamma$ fit method to a set of realistic models of HD 69686
to investigate more closely the difference between the statistically estimated
$\log g_{\rm polar}$ and the true values of $\log g_{\rm polar}$ for models
at various inclination angles.  The model star has a mass of 4.42 $M_\odot$.
At different inclination angles, we synthesize the H$\gamma$ profiles by
considering Roche geometry (controled by both rotational and gravitational
potential on the surface) and the gravity darkening
effect ($T_{\rm local} \propto g_{\rm eff}^{0.25}$).  We adjust $T_{\rm polar}$
and $R_{\rm polar}$ of the model stars for each inclination angle so that
the measured values ($T_{\rm msr}$ and $\log g_{\rm msr}$ from H$\gamma$ fit)
always equal the values given in Table~2.  All results
are given in Table~4, where we can see that the range of $\log g_{\rm polar}$,
$4.04\pm 0.04$, covers all inclination angles from $i=90^\circ$ down to
$i=30^\circ$.  Below $i=30^\circ$ (i.e., $V_{\rm eq} > 282$ km~s$^{-1}$),
the model star spins close to the critical speed (around $i=20^\circ$).
Because of low probability that the star rotates with higher $V_{\rm eq}$ ($> 280$ km~s$^{-1}$)
\citep{hua06a} and because the corresponding inclination range is narrow
($20^\circ < i < 30^\circ$), the chance of HD 69686 spinning at a low inclination
angle and  having a higher
$\log g_{\rm polar}$ beyond the range of $4.04\pm 0.04$ is low.  Even if HD 69686
does spin at a low inclination angle, its age would be older than that its polar
gravity implies because it rotates much faster than at a high inclination angle
(see Table~3).  In this numerical simulation, we have
also checked the projected area of the models on sky.  We found that the
projected area of the models at all investigated inclination angles are
basically same (different within $\pm1$\%).  As the inclination angle goes
lower, the $f$ factor gets bigger.  Meanwhile the model star rotates faster
and faster (increasing $V_{\rm eq}$). In order to keep the measured
$\log g$ constant, we have to lower the polar radius, $R_{\rm polar}$.
Thus, the net effect is that the projected area on the sky, $\pi R_{\rm polar}^2/D^2$
multiplied by $f$, stays almost unchanged.  In other words, it is safe to derive
the distance using the projected area at one inclination angle.

\placetable{tab4}      

It should be noted that, as we were deriving the surface $\log g$
of HD 69686 from H$\gamma$ profile fit, we did not consider possible
systematic errors in the synthesized profiles.  Our synthesis code
took the specific intensity profile data generated by
SYNSPEC43\footnote{http://nova.astro.umd.edu/Synspec43/synspec.html}, which
is using the VCS theory \citep{vid73}.  If the systematic error causes
an underestimation of surface gravity, $\log g$, by 0.1 dex or more,
HD 69686 might be younger than we estimate above, and could be
born on the Galactic plane (see section 4).

Many previous studies \citep{tob81,saf97,mag01,ram01a,lyn04,mar04} have pointed
out that some old evolved objects in the halo, lying on a post-horizontal
branch (PHB) or a post-asymptotic giant branch (PAGB), could have similar
atmospheric parameters (temperature and gravity) as young massive OB stars
of population I.  Could HD~69686 be an old PHB or PAGB halo object with
the same $T_{\rm eff}$ and $\log g$ values listed in Table~2?  One 
key difference between the old blue halo objects and the young OB disk
stars is that the former have much lower mass ($<$ 1$M_\odot$).  If
HD~69686 had a mass lower than $1 M_\odot$,  then for its observed
gravity, it would have a small radius and a distance of less than 180~pc,
which would yield a parallax more than 5$\sigma$ higher than the Hipparcos
parallax ($2.35\pm0.58$ mas).  Thus, we are confident that HD~69686 is a
young, massive, main sequence B star instead of an old blue halo star with
similar atmospheric parameters.


\section{Tracing Backward to the Birthplace}                   

Now we have very accurate information that we list in Table~2
about the position and the space velocity of HD~69686 
(resolved into three orthogonal components relative to the Sun: 
the radial velocity $V_r$ and the tangential velocity in 
the directions of increasing right ascension $V_{\rm RA}$ 
and declination $V_{\rm Dec}$ for epoch J2000). 
With help of a properly selected Galactic potential model, we should
be able to follow the motion of HD~69686 backward to its birthplace.
Among the many Galactic models available in literature, we decided to
choose the axisymmetric model proposed by \citet{deh98b} (their model 2,
DB2 here after).  Their model mass distribution was constructed to meet
the constraints imposed by all the high quality kinematical data that
were available at that time.  The gravitational acceleration is related
to the potential by $\overrightarrow{a}(R,z)=-\nabla \Phi(R,z)$.  We
obtained both the potential and the acceleration data of the model in
a format of 3D grids in space (0.1 kpc per step in $R$ and 0.01 kpc
per step in $z$), and we made a bilinear-interpolation in this grid to
determine the acceleration at each time step.

The star's true space velocity in the Galactic potential field
is its velocity relative to the Local Standard
of Rest (LSR, hereafter), ($U$,$V$,$W$), plus the Galactic circular velocity
of LSR, (0, $V_{\rm circ}(R_0)$,0).
Here $R_0$ ($=8$ kpc) is the Sun's distance to the Galactic center \citep{rei93},
and $V_{\rm circ}(R_0)=$217 km~s$^{-1}$ is the circular velocity of DB2
at $R_0$ (very close to the IAU recommended value, 220 km~s$^{-1}$).

In order to trace the star's motion back in time, 
we use the reversed space velocity of HD~69686,
$-$($U$,$V+V_{\rm circ}(R_0)$,$W$) as the initial velocity.  The initial
location of HD~69686 in the Galaxy is determined from its location relative
to the Sun and the Galactic location of the Sun, 8 kpc away from the Galactic
center \citep{rei93} and 8 pc above the Galactic plane \citep{hol97}.  We calculated
$\overrightarrow{a}(R,z)$ first, and then moved the star a very small step
assuming $\overrightarrow{a}(R,z)$ stays constant during this small move.
We updated the star's velocity and acceleration at the end of each step.  Repeating
this calculation until reaching the estimated age of the star (actually
reaching to the zero age of the star because we move back in time),
the integration follows the star back to its birthplace.  The numerical
errors of this integration depend on the size of each small step.  We
found that a step of 2 kyr is a good choice which results in only a very
small difference ($<$ 1 pc) from the integrated positions using a
smaller step (1 kyr) over 80 Myr.

\begin{figure}
\plotone{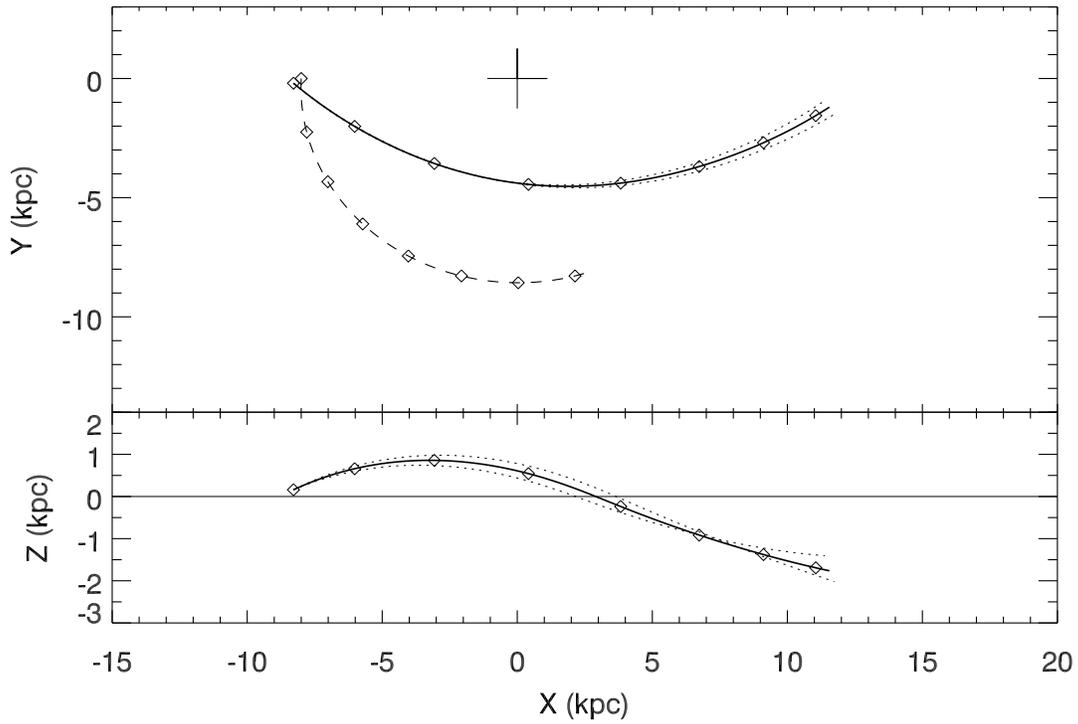}
\caption{The trajectory of HD~69686 (solid line) during the past 73 Myr
based on the Galactic model 2 by \citet{deh98b} ($V_{\rm circ}=217$ km~s$^{-1}$ at $d=$ 8 kpc).
The trajectories with slightly different initial velocities (by 1 $\sigma$ off in $V_{\rm RA}$)
are plotted as dotted lines.  The orbit of the Sun is
plotted as a dashed line in the upper panel.  The Galactic center is marked by the big plus
sign.  Diamonds indicate time intervals of 10 Myr.
}
\label{fig4}
\end{figure}

\placefigure{fig4}     

The reconstructed trajectory of HD~69686
is plotted in Figure~4 as a thick solid line.  The dotted lines are
those tracks that result by changing by 1~$\sigma$ the space velocity 
component $V_{\rm RA}$ (with the dominant velocity error, $1 \sigma=8$ km~s$^{-1}$).
Because of the very large peculiar velocity of HD~69686, it has a non-circular
orbital motion (the circular orbit of the Sun in the $X-Y$
plane is plotted as a dashed line in Fig.~4 for comparison).  Currently,
HD~69686 is quickly passing through the solar neighborhood and moving
away from the Galactic center ($U=-191$ km~s$^{-1}$).  When we first derived
$U$, we suspected that this star might come from the very inner part of
the Galaxy.  After the full trajectory was reconstructed, we realized
that HD~69686 has traveled a very long distance across the Galaxy before
reaching its current location (see the top panel of Fig.~4).  
The birthplace of HD~69686 is located in the outer part of the Galaxy
(about 12 kpc from the Galactic center, 4 kpc further out than the Sun).
At its birth (73 Myr ago), HD~69686 was more than 10~kpc away from the
Sun (compared to its current distance of 380~pc).  About 26.6 Myr ago,
HD~69686 reached its minimum distance from the Galactic center
($\approx 4.4$ kpc).

\citet{mar06} studied more than 60 high Galactic latitude B stars and he suggested
that all the young population I B stars in his sample are likely ejected from the
Galactic plane.  \citet{mag01} and \citet{ram01b} found very similar results 
in their investigations of high-latitude blue objects.  
If the cause of the high peculiar velocity of
HD~69686 is similar to other runaway OB stars, then its birthplace is expected
to be in or very close to the Galactic plane because star formation activities
are very rare at places far from the plane.  However, our calculated
trajectory suggests that HD~69686 is an exception.  In the lower panel of Figure~4,
HD~69686's projected trajectory on the $X-Z$ plane clearly shows that it was
located well below the Galactic plane ($z \sim -1.8$ kpc) 73 Myr ago.
Today HD~69686 (only 150 pc above the Galactic plane) is on its way plunging
into the Galactic plane ($W=-57$ km~s$^{-1}$).  Moving backward in time,
it reached the highest position of its trajectory (860 pc above the
Galactic plane) 19.5 Myr ago, and crossed the plane around
37 Myr ago, thanks to the stronger gravitational force in the $z$ direction
in the inner part of the Galaxy.  However, further back in time, 
HD~69686 was below the Galactic plane and further away from the
Galactic center, and, therefore, it experienced a weaker $z$-force.
The $z$-force was so weak that the star's trajectory began
73 Myr ago at a position well below the plane ($z \sim -1.8$ kpc) 
and with a velocity approaching the plane 
($W=+22$ km~s$^{-1}$).  The errors in the velocity components and
the age estimation of HD~69686 are not large enough to make the range
of its possible birthplace extend to the Galactic plane.

\placefigure{fig4}     


\section{Discussion}                             

The reconstructed trajectory of HD~69686 implies that HD~69686
might be born in a location with a particularly low density (1.8 kpc
below the Galactic plane and 12 kpc away from the Galactic center),
where star formation activity is expected to be very rare.  Of course,
we assume here that HD~69686 is a single-star system.
Though our two consecutive night observations do not display any variation
in radial velocity above the noise level,  we can not completely
rule out that HD~69686 is a binary system with a longer
period ($>10$ days).  The mass-transfer events that often occur
in close binary systems could totally invalidate our age estimation
based on single-star evolutionary tracks.  If HD~69686 is in a binary
system and was ejected from the Galactic plane due to the supernova (SN)
explosion of its companion, the explosion occured most likely about
37 Myr ago when HD 69686 was crossing the Galactic plane.  The coordinates
of the crossing position are $(x,y,z)=(2919,-4477,0)$ pc, and the space
velocity of HD~69686 at that moment is
$(V_{\rm x},V_{\rm y},V_{\rm z})=(-326.2, -24.4, 84)$ km~s$^{-1}$
while the Galactic circular velocity at that location is
$(-181.8, -118.5, 0)$ km~s$^{-1}$ (using DB2).  Then we can calculate
HD 69686's peculiar velocity (i.e.\ ejection velocity) at that location,
which is 192 km~s$^{-1}$.  \citet{nel99} investigated the relation
between the ejection velocity, the mass loss during the SN explosion,
and the current binary period.  For HD 69686 with a long period ($> 10$ days)
and a very high ejection velocity (192 km~s$^{-1}$),  the equation (7)
in \citet{nel99} suggests a very large SN mass loss ($> 10 M_\odot$),
which is large for a binary containing a late B-type star.  Although the over-solar abundance
of helium on HD 69686's surface could be a result of past mass
transfer in a close binary, it is not the only explanation because
helium peculiar (both rich and poor) stars are quite common among young
late B (single) stars \citep{hua06b}. Also, as described in following
paragraphs, we found evidence indicating that HD 69686 seems to belong
to a co-moving cluster that naturally explains its high peculiar
velocity.  All of these arguments suggest that HD 69686 does not
belong to a binary system, and we can estimate
its age using single star evolutionary tracks.

Much effort has been invested in searching for young
massive stars forming in situ in the halo \citep{zel97,lyn02,lyn04,mar04,
mar06}.  Most of the blue halo objects examined in these studies are
either ejected disk stars or old evolved stars (PHB or PAGB).
Only a few stars are possible candidates of massive
star formation in the halo.  \citet[and references therein]{del08} 
summarized the observational evidence of star formation far from
the Galactic plane.  One possible mechanism that can trigger star
formation at such an unusual place is a collision between high velocity
clouds (HVC) and/or Galactic accretion fragments (or tidal streams).
\citet{dys83} proposed that early-type star formation can be triggered
by ``cloudlet-cloudlet collisions occurring every $10^8$ yr'' in
any particular high Galactic latitude cloud, and the formation rate
could be $10^3$ early-type stars every $10^8$ yr in the halo.
Later, \citet{chr97} suggested that the collisions between cloudlets
within HVCs may occur less frequently than previously thought, and they
derive a much lower star formation rate from the observational data.
More recently, another mechanism of star formation in the halo was
proposed by \citet{del08}, i.e., that the tidal force of a passing
massive cluster (such as a globular cluster) can trigger star formation
in high Galactic latitude clouds.  Though the proposed mechanisms
are different, they all suggest a common scenario that a group of
stars should be formed together in situ in the halo.  If this indeed
is the case for HD~69686, we expect to see a group of stars that share
the same space velocity with and have the same age of HD~69686.
Finding such co-moving stars would provide strong evidence to support
their formation in situ in the halo.

Because HD~69686 has traveled a very long distance in 73 Myr, it is
expected that the co-forming stars, if they exist, would spread into a much
larger space around the current position of HD~69686 due to the velocity
dispersion in the original cluster.  A dispersion of 10 km~s$^{-1}$ in
original space velocity would cause the co-forming group spread into
a volume of 750$^3$ pc$^3$, i.e., spread all over
the sky as seen from the Sun.  To make the search viable,
we limit our search range to the vicinity of the trajectory of HD~69686.
Thus, our search will detect only a very small fraction of any co-moving
stars whose space velocities are similar to that of HD~69686.

\begin{figure}
\plotone{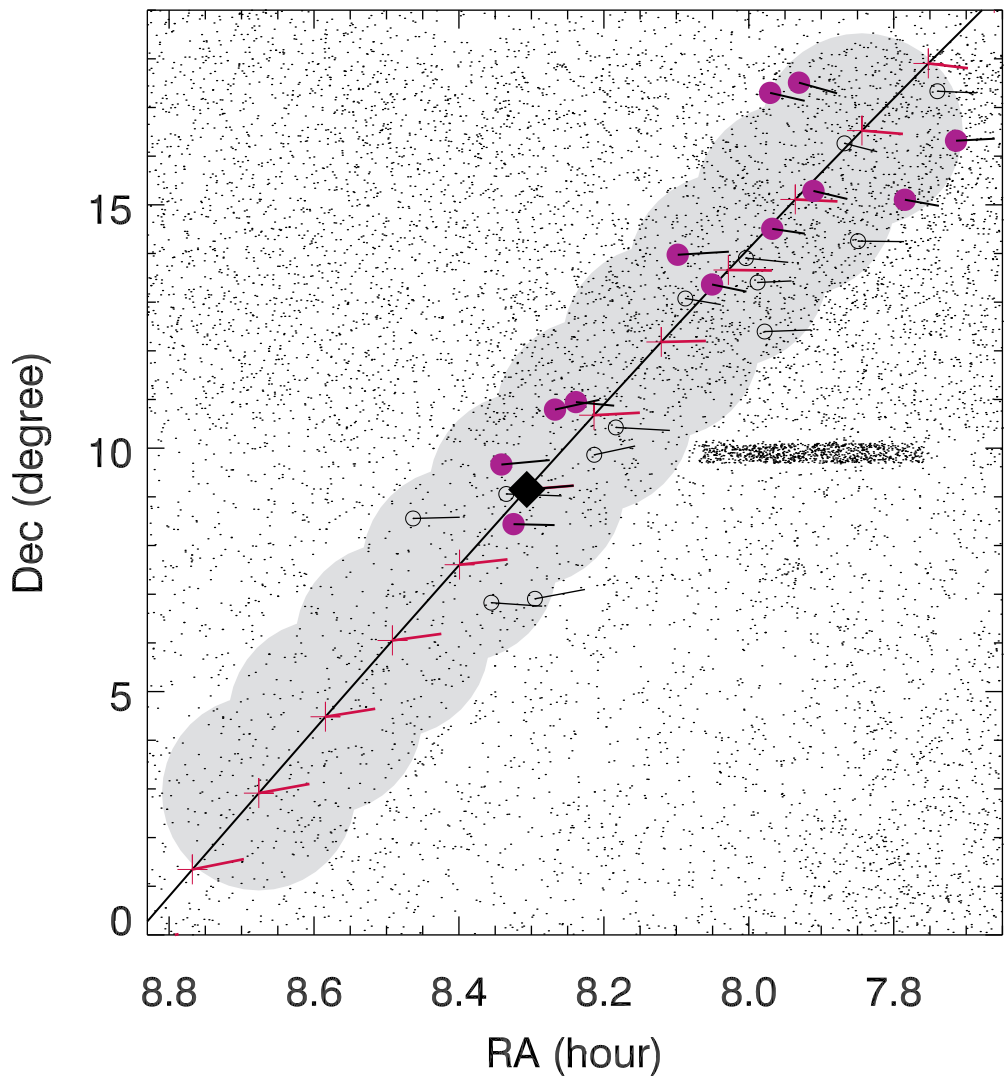}
\caption{The sky map of high proper motion stars around HD~69686.  HD~69686 is
plotted as a diamond.  The candidate members of the HD~69686 co-moving
group are plotted as filled circles. The foreground/background stars with similar
proper motions are plotted as open circles.  All other high proper motion stars
from the Catalogue of Stars with High-Proper Motions (V2) \citep{iva08} are plotted
as small dots.  The solid line is the projected trajectory of HD~69686 on the sky.
The plus signs mark the positions at 0.05 Myr intervals.  The 
proper motion lines associated with individual
objects indicate the size of their motions on the sky over a time period of 0.04 Myr.
}
\label{fig5}
\end{figure}

\placetable{tab5}      
\placefigure{fig5}     
 
To carry out this kind of search, we first need the projected trajectory
of HD~69686 on the sky as a baseline.  Any co-moving objects
are expected to be on or around this baseline and to have similar proper
motions and radial velocities to the values of the nearest point on the
trajectory.  We calculated the projected positions of HD~69686's trajectory
on the sky every 0.05 Myr between 0.5 Myr in the past and 0.5 Myr in the
future.  The results are given in Table~5 and plotted in Figure~5.  We
used the Catalogue of Stars with High-Proper Motions 
(V2)\footnote{http://cdsarc.u-strasbg.fr/viz-bin/Cat?I/306A}
by \citet{iva08} to search for similar high proper motion stars.
The search region (about 2 degrees around the trajectory) is
indicated as a shadowed area in Figure~5.  We found a total of 26 stars
(including HD~69686) from roughly 2000 high proper motion stars
($> 40$ mas~yr$^{-1}$) in the search region.  Their proper motions (in both RA and
Dec directions) have differences of 15 mas~yr$^{-1}$ or less from the
reference points on the trajectory of HD~69686.  All 25 stars other
than HD~69686 are plotted in Figure~5 as open or filled circles with their proper
motions indicated by line segments.  

Note that the proper motions illustrated in Figure~5 are not
aligned with the projected trajectory (HD~69686 seems to move away from
its own trajectory).  The trajectory of HD~69686 in space (Fig.~4)
is a curve in the Galactic potential field that differs from circular
motion.  In order to find the co-moving stars that move on a similar
trajectory, we need to project the whole trajectory on the current sky.
This means that both the past and future parts of the trajectory
of HD~69686 have to be projected on the present sky.  However,
HD~69686's proper motion is related to projected positions on the
sky at different moments, i.e., the past-time positions on the
past-time sky and the future positions on the future sky.  This
is different from their position in the present sky (the solid
line in Fig.~5) because of the Sun's orbital motion in the Galaxy.
As the Sun (and Earth) is moving at $\sim$220 km~s$^{-1}$ around the
Galactic center, the projected trajectory of HD~69686 (the solid line)
will move downward (and slightly to the left) in Figure~5 (in the reverse
direction of the solar circular motion, which is approximately parallel
to Galactic latitude).  Thus, although HD~69686 seems to move away from
its projected trajectory on the sky, it actually stays on the trajectory
because the projected trajectory is moving downward at the same time.

\begin{figure}
\plotone{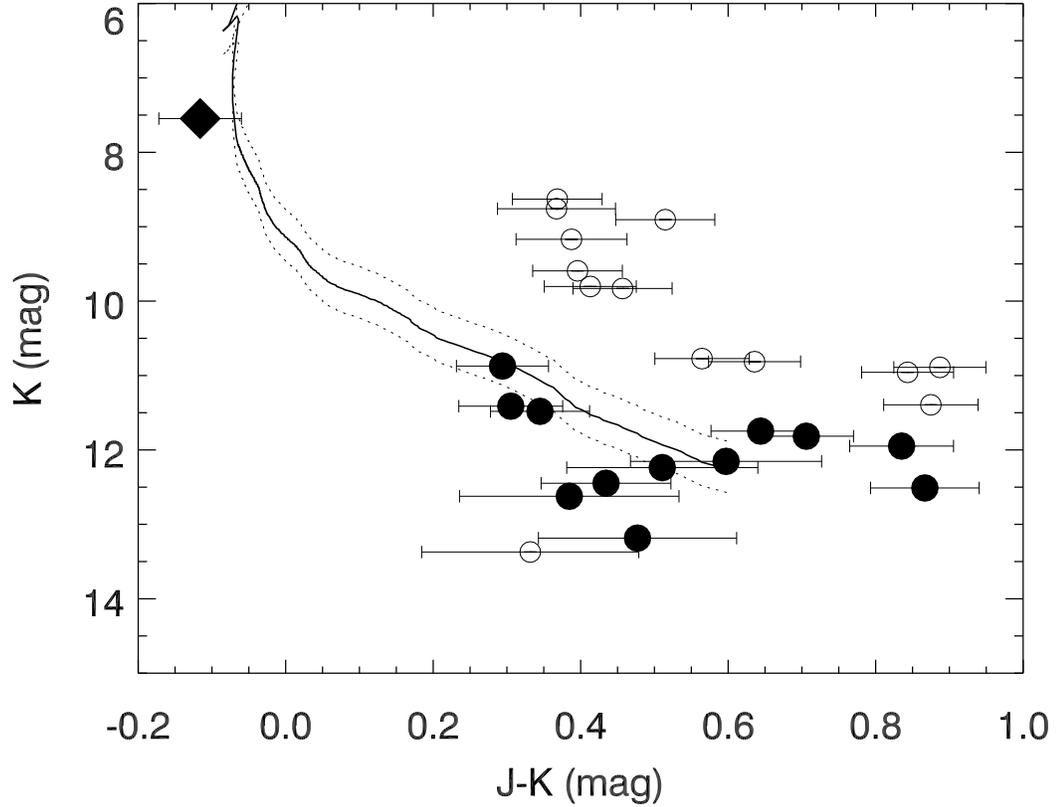}
\caption{The color-magnitude diagram of the HD~69686 co-moving group. HD~69686 is
plotted as a diamond.  The candidate members of the HD~69686 co-moving
group are plotted as filled circles. The foreground/background stars with similar
proper motions are plotted as open circles.  The solid line is an isochrone
of log Age = 7.84 and metallicity $Z=0.020$ from \citet{lej01} for a distance
of 380 pc. The dotted lines are the same isochrone at distances 380$\pm$60 pc.
}
\label{fig6}
\end{figure}

\placefigure{fig6}     
\placetable{tab6}      

Having similar proper motions does not necessarily imply that
these stars belong to a co-moving group.  Some foreground and
background high proper motion stars can be mixed in.  We can
further sift these foreground/background stars out using the
color-magnitude diagram (CMD) of the expected co-moving group.
All members in the HD~69686 co-moving group are expected to have
the same age ($\sim$73 Myr) and a distance around 380 pc (because
we limit our search region in the vicinity of the trajectory whose
distance is given in the last column of Table~5).  A 70 Myr
isochrone (for solar metallicity $Z=0.02$ and $\log$ Age $=7.84$)
extracted from evolutionary models by \citet{lej01} is plotted
in Figure~6 assuming a distance of 380 pc and $A_V=0$.  
The $JHK_s$ magnitudes of all 26 stars were obtained from the 2MASS
catalogue \citep{skr06} and were transformed into the system of
\citet{bes88} using formulae\footnote{www.astro.caltech.edu/$^\sim$jmc/2mass/v3/transformations/}, $(J-K)_{\rm BB}=[(J-K)_{\rm 2M}+0.018]/0.983$ and
$K_{\rm BB}=(K_s)_{\rm 2m}-0.001(J-K)_{\rm BB}+0.039$.
From Figure~6, we can easily tell those foreground/background
stars which fall far above or below the isochrone on the CMD.
This step helped us to discard 13 out of 25 stars (the 13 deleted
objects are plotted as open circles in Fig.~5 and Fig.~6).  The remaining
12 stars plus HD 69686 are the final candidates of the high velocity
co-moving group (plotted as filled circles in Fig.~5 and Fig.~6).
The key parameters of the final candidates are listed in
Table~6.   However, we cannot be sure that 
all these 12 stars co-move with HD~69686 because
we do not have the radial velocity data for them yet.  New
spectroscopic observations of these stars are planned in the
near future.  If some of the 12 stars are confirmed to co-move
with HD~69686 by the follow up observation in the future, then
we can rule out a runaway origin for HD~69686 
because the mechanisms of generating runaways
cannot account for the ejection of a group of stars in a common direction.
Instead, the existence of a co-moving group would support the idea that 
the massive star formed in situ in the halo
(most likely in a HVC or Galactic merger fragment) with other
stars 73 Myr ago.

The trajectory of HD~69686 covers a large range in all dimensions of
the Galactic potential field ($4\sim 12$ kpc in $R$, $-1.8\sim +0.9$ kpc in $z$,
and almost $180^\circ$ of Galactic azimuth).  Its shape
may depend on the selection of different Galactic models.  We did calculate
the trajectory using different models \citep{pac90,all91} and found
similar results with trivial differences that are too small
to change our analysis results.  All the models used in our calculations
assume a similar circular velocity at a distance to Galactic center of 
8~kpc ($\sim 220$ km s$^{-1}$).  We note that some observations
\citep{miy98, men00, uem00, sha09, rei09} suggest that the circular
motion of the solar neighborhood is faster than 220 km s$^{-1}$
by 30$\sim$40 km s$^{-1}$.  Although it will be very interesting to see
how much the trajectory of HD~69686 would change in a Galactic model
characterized by a larger rotation speed at the solar circle, it is
beyond the scope of the present paper.


\acknowledgments
We thank our referee, Dr Philip Dufton, very much for his careful
reading and constructive comments on our paper.
We thank Dianne Harmer and the KPNO staff for their assistance
in making these observations successful.  This material is
based upon work supported by the National Science Foundation 
under Grant No.~AST-0606861.  WJH thanks Dr.\ George Wallerstein and
the Kenilworth Fund of the New York Community Trust for partial financial
support of this study.  MVM is grateful for support from Lehigh University.
This publication makes use of data products from the Two Micron
All Sky Survey, which is a joint project of the University of
Massachusetts and IPAC/Caltech, funded by NASA and NSF.



\clearpage
\begin{deluxetable}{lcc}
\tablewidth{0pc}
\tablecaption{Radial Velocity Measurements
\label{tab1}}
\tablehead{
\colhead{ } &
\colhead{$V_r$(N1)\tablenotemark{a} } &
\colhead{$V_r$(N2)\tablenotemark{b} } \\
\colhead{Lines } &
\colhead{(km~s$^{-1}$) } &
\colhead{(km~s$^{-1}$) }}
\startdata
\ion{He}{1}~$\lambda$4471  &  149.0   &  148.8  \\
\ion{Mg}{2}~$\lambda$4481  &  145.0   &  148.2  \\
\ion{H}{1}~$\lambda$4340  &   149.2   &  145.0  \\
\enddata
\tablenotetext{a}{HJD(N1)$=$2,454,788.0079}
\tablenotetext{b}{HJD(N2)$=$2,454,789.0227}
\end{deluxetable}

\clearpage

\begin{deluxetable}{lc}
\tablewidth{0pc}
\tablecaption{Parameters of HD~69686
\label{tab2}}
\tablehead{
\colhead{Parameter} &
\colhead{Value }}
\startdata

$l$ (deg)                   &           214.5041 \\
$b$ (deg)                   &           23.4482 \\
$V_{\rm r}$ (km~s$^{-1}$)   &          148$\pm$6  \\
$V_{\rm RA}$ (km~s$^{-1}$)  &          $-$155$\pm$8 \\
$V_{\rm Dec}$ (km~s$^{-1}$) &           13$\pm$1  \\
$V \sin i$ (km~s$^{-1}$)    &           141$\pm$7  \\
$T_{\rm eff}$ (K)           &          14760$\pm$200  \\
$\log g$ (dex)              &          3.93$\pm$0.03  \\
$\log g_{\rm polar}$ (dex)  &          4.04$\pm$0.04  \\
Mass ($M_\odot$)            &          4.42$\pm$0.15 \\
Age (Myr)                   &          73$\pm$10  \\
Distance (pc)               &          380$\pm$20  \\
                            &                      \\
$U_{\rm LSR}$\tablenotemark{a} (km~s$^{-1}$) &          $-191$  \\
$V_{\rm LSR}$\tablenotemark{a} (km~s$^{-1}$) &          $-37$   \\
$W_{\rm LSR}$\tablenotemark{a} (km~s$^{-1}$) &          $-57$   \\
\enddata
\tablenotetext{a}{The peculiar velocity of the Sun
is $U_0=10.0$, $V_0=5.25$, and $W_0=7.17$ \citep{deh98a}.}
\end{deluxetable}
\clearpage

\begin{deluxetable}{cc}
\tablewidth{0pc}
\tablecaption{Evolution Time of Rotating Stars\tablenotemark{a}
\label{tab3}}
\tablehead{
\colhead{$\Omega/\Omega_{\rm crit}$ } &
\colhead{Evolving Time\tablenotemark{b}(Myr)}}
\startdata

0.1 &   244  \\
0.3 &   262  \\
0.5 &   269  \\
0.7 &   275  \\
0.8 &   279  \\
0.9 &   281  \\

\enddata
\tablenotetext{a}{This table is based on the data of the rotating models of
3 $M_\odot$ from \citet{eks08}.}
\tablenotetext{b}{The time that a rotating model takes to evolve from
ZAMS to $\log g_{\rm polar} = 4.04$.}
\end{deluxetable}

\clearpage

\begin{deluxetable}{cccc}
\tablewidth{0pc}
\tablecaption{Realistic Stellar Models at Different Inclination Angles\tablenotemark{a}
\label{tab4}}
\tablehead{
\colhead{Inclination} &
\colhead{$T_{\rm polar}$ } &
\colhead{$R_{\rm polar}$ } &
\colhead{$\log g_{\rm polar}$ } \\
\colhead{(degree)} &
\colhead{(K) } &
\colhead{($R_\odot$) } &
\colhead{(dex)}}
\startdata

90 &   15260  &  3.52  &  3.99  \\
80 &   15270  &  3.52  &  3.99  \\
70 &   15310  &  3.50  &  4.00  \\
60 &   15370  &  3.48  &  4.00  \\
50 &   15440  &  3.43  &  4.01  \\
40 &   15525  &  3.32  &  4.04  \\
30 &   16260  &  3.13  &  4.09  \\
25 &   16960  &  2.95  &  4.14  \\

\enddata
\tablenotetext{a}{All models have the same stellar mass ($=4.42 M_\odot$) and
same $V \sin i$ ($=141$ km~s$^{-1}$).  The parameters of the models in this table
are prepared in such way that, when we apply our H$\gamma$ profile fit
method to the synthesized H$\gamma$ profiles of these models, we obtain
the same $T_{\rm msr}$ ($=14760$K) and $\log g_{\rm msr}$ ($=3.93$).}
\end{deluxetable}

\clearpage

\begin{deluxetable}{rccrrcc}
\tablewidth{0pc}
\tablecaption{Projected Trajectory of HD~69686 on the Sky
\label{tab5}}
\tablehead{
\colhead{Time} &
\colhead{RA } &
\colhead{Dec } &
\colhead{$\mu_\alpha \cos \delta$ } &
\colhead{$\mu_\delta$ } &
\colhead{$V_r$ } &
\colhead{Dist. } \\
\colhead{(Myr)} &
\colhead{(J2000) } &
\colhead{(J2000) } &
\colhead{(mas/yr) } &
\colhead{(mas/yr) } &
\colhead{(km~s$^{-1}$) } &
\colhead{(pc)}}
\startdata

$-$0.50 &  09 12 58.4 & $-$06 18 28 &  $-$101.0 &    26.7  &   100   &   389  \\
$-$0.45 &  09 07 41.2 & $-$04 49 00 &  $-$100.3 &    25.6  &   106   &   386  \\
$-$0.40 &  09 02 21.0 & $-$03 18 07 &   $-$99.5 &    24.3  &   111   &   384  \\
$-$0.35 &  08 56 57.9 & $-$01 46 02 &   $-$98.5 &    22.7  &   116   &   381  \\
$-$0.30 &  08 51 32.2 & $-$00 13 00 &   $-$97.3 &    21.0  &   121   &   380  \\
$-$0.25 &  08 46 04.3 & $+$01 20 41 &   $-$95.9 &    19.1  &   126   &   378  \\
$-$0.20 &  08 40 34.5 & $+$02 54 47 &   $-$94.3 &    17.0  &   131   &   378  \\
$-$0.15 &  08 35 03.2 & $+$04 28 57 &   $-$92.5 &    14.7  &   135   &   377  \\
$-$0.10 &  08 29 30.5 & $+$06 02 56 &   $-$90.6 &    12.3  &   140   &   378  \\
$-$0.05 &  08 23 57.0 & $+$07 36 24 &   $-$88.5 &     9.8  &   144   &   378  \\
 0.00 &  08 18 23.0 &   $+$09 09 05 &   $-$86.2 &     7.2  &   148   &   380  \\
 0.05 &  08 12 48.8 &   $+$10 40 41 &   $-$83.8 &     4.6  &   152   &   381  \\
 0.10 &  08 07 14.9 &   $+$12 10 59 &   $-$81.3 &     1.8  &   155   &   383  \\
 0.15 &  08 01 41.5 &   $+$13 39 41 &   $-$78.7 &    $-$0.9 &    158  &    386  \\
 0.20 &  07 56 09.0 &   $+$15 06 37 &   $-$75.9 &    $-$3.6 &    161  &    389  \\
 0.25 &  07 50 37.8 &   $+$16 31 33 &   $-$73.2 &    $-$6.2 &    164  &    392  \\
 0.30 &  07 45 08.3 &   $+$17 54 19 &   $-$70.3 &    $-$8.8 &    166  &    396  \\
 0.35 &  07 39 40.7 &   $+$19 14 46 &   $-$67.4 &   $-$11.4 &    169  &    401  \\
 0.40 &  07 34 15.4 &   $+$20 32 48 &   $-$64.5 &   $-$13.8 &    171  &    405  \\
 0.45 &  07 28 52.7 &   $+$21 48 19 &   $-$61.6 &   $-$16.1 &    172  &    411  \\
 0.50 &  07 23 32.9 &   $+$23 01 14 &   $-$58.8 &   $-$18.4 &    174  &    416  \\
\enddata
\end{deluxetable}
\clearpage

\begin{deluxetable}{ccrrrrccc}
\rotate
\tablewidth{0pc}
\tablecaption{Candidates of the HD~69686 Co-Moving Group\tablenotemark{a}
\label{tab6}}
\tablehead{
\colhead{RA } &
\colhead{Dec } &
\colhead{$\mu_\alpha \cos\delta$ } &
\colhead{$\Delta\mu_\alpha \cos\delta$ } &
\colhead{$\mu_\delta$ } &
\colhead{$\Delta\mu_\delta$} &
\colhead{} &
\colhead{$J$ } &
\colhead{$K_s$ } \\
\colhead{(J2000) } &
\colhead{(J2000) } &
\colhead{(mas/yr) } &
\colhead{(mas/yr) } &
\colhead{(mas/yr) } &
\colhead{(mas/yr) } &
\colhead{Ref.\tablenotemark{b}}  &
\colhead{(mag) } &
\colhead{(mag)}}
\startdata
08 18 22.98 & +09 09 05.1 &  $-$86.2 & 0.67 &  7.2  & 0.42 & 1 & 7.38 &   7.51 \\
08 20 29.24 & +09 39 52.6 &  $-$88.7 &      5.9 &   8.1 &     5.6 & 2 & 13.60 &  13.15 \\
08 19 28.74 & +08 26 23.8 &  $-$79.4 &  \nodata &   3.0 & \nodata & 3 & 11.66 &  11.37 \\
08 14 18.90 & +10 56 48.2 &  $-$70.0 &     36.0 &  $-$6.6 &    37.5 & 4 & 12.82 &  12.41 \\
08 16 02.71 & +10 47 35.2 &  $-$81.4 &     28.9 &  18.1 &    30.5 & 4 & 12.68 &  12.12 \\
08 05 53.09 & +13 58 28.0 &  $-$93.2 &  \nodata &   5.1 & \nodata & 3 & 10.90 &  10.83 \\
08 03 02.59 & +13 21 47.1 &  $-$62.4 &      5.5 & $-$13.2 &     5.4 & 2 & 12.94 &  12.58 \\
07 54 38.84 & +15 17 09.0 &  $-$60.9 &      2.1 & $-$14.9 &     2.0 & 2 & 11.76 &  11.44 \\
07 58 03.59 & +14 30 24.8 &  $-$58.8 &      5.2 &  $-$8.9 &     5.2 & 4 & 12.45 &  11.78 \\
07 55 50.86 & +17 30 35.4 &  $-$67.8 &      5.8 & $-$19.7 &     5.8 & 4 & 13.31 &  12.47 \\
07 47 04.61 & +15 06 15.7 &  $-$61.3 &      6.0 & $-$11.2 &     5.9 & 4 & 12.71 &  11.91 \\
07 42 50.37 & +16 19 00.7 &  $-$69.1 &      6.1 &   3.8 &     6.0 & 2 & 13.01 &  12.20 \\
07 58 14.04 & +17 18 01.8 &  $-$61.5 &     29.5 & $-$14.8 &    31.0 & 4 & 11.78 &  11.71 \\
\enddata
\tablenotetext{a}{The first row is for HD~69686.}
\tablenotetext{b}{The Catalogue of Stars with High Proper Motions \citep{iva08} does
 not provide the errors in the proper motions.
 This column indicates sources for the errors: 1 - \citet{van07}, 2 - \citet{zac04}, 
 3 - \citet{kis00}, 4 - \citet{duc06}.} 
\end{deluxetable}

\clearpage

\clearpage

\clearpage

\clearpage

\clearpage

\clearpage

\end{document}